\documentclass[12pt,english]{article}
\usepackage[T1]{fontenc}
\usepackage[latin1]{inputenc}
\usepackage{a4wide}
\usepackage{geometry}
\usepackage{babel}
\usepackage{graphics}

\makeatletter

\providecommand{\LyX}{L\kern-.1667em\lower.25em\hbox{Y}\kern-.125emX\@}
\newcommand{\noun}[1]{\textsc{#1}}

 \newcommand{\lyxaddress}[1]{
   \par {\raggedright #1 
   \vspace{1.4em}
   \noindent\par}
 }

\usepackage{amssymb}
\makeatother

\begin{document}

\title{Boundary reduction formula}

\author{Z. Bajnok\protect\( ^{\dagger }\protect \), G. Böhm\protect\( ^{*}\protect \)
and G. Takács\protect\( ^{\dagger }\protect \)}

\maketitle

\lyxaddress{\centering \emph{\protect\( ^{\dagger }\protect \)Institute for Theoretical
Physics Eötvös University }\\
\emph{H-1117 Budapest Pázmány s. 1/A}\\
\emph{\protect\( ^{*}\protect \)Research Institute for Nuclear and Particle
Physics}\\
\emph{H-1525 Budapest, 114 P.O.B. 49, Hungary}}

\begin{abstract}
An asymptotic theory is developed for general non-integrable boundary quantum
field theory in 1+1 dimensions based on the Langrangean description. Reflection
matrices are defined to connect asymptotic states and are shown to be related
to the Green functions via the boundary reduction formula derived. The definition
of the \( R \)-matrix for integrable theories due to Ghoshal and Zamolodchikov
and the one used in the perturbative approaches are shown to be related.
\end{abstract}

\subsection*{Introduction}

Two dimensional boundary quantum field theories have been analysed from two
different points of view, the bootstrap and the perturbative, respectively.

The former was initiated by Ghoshal and Zamolodchikov in \cite{GZ} and can
be applied to integrable theories. In such theories there is an infinite number
of conserved quantities, which give severe restrictions on the allowed physical
processes: Besides the usual constraints such as factorization and purely elastic
bulk scattering there is also factorization and purely elastic reflection on
the boundary. The scattering theory developed in \cite{GZ} is analogous to
axiomatic scattering theory \cite{ELOP}: in the \emph{in} state the particles
travel towards the boundary with decreasingly ordered momenta, while in the
\emph{out} state, where all the scatterings and reflections have been terminated,
they travel away from the boundary with decreasingly ordered momenta again.
The \( R \)-matrix which connects the \emph{in} and \emph{out} states is the
composition of the individual reflection and the pairwise scattering matrices.
The one particle reflection matrices have to obey unitarity, boundary Yang-Baxter
and boundary crossing relations. Using these relations together with the bootstrap
condition (\cite{GZ},\cite{FK}) the model can be solved modulo CDD type ambiguities.
We emphasise that in the bulk case the axioms of the scattering theory such
as unitarity, crossing symmetry \cite{ELOP} were motivated by relativistic
field theoretic results based on the perturbative, Lagrangean description. In
the boundary case, however, to our knowledge, no such background is available.
In \cite{GLM} the non-linear Schrödinger model with linear boundary condition
on the half line is considered and the assumptions of the axiomatic scattering
theory are rigorously checked. This model is, however, non-relativistic and
integrable. 

The perturbative approach to boundary quantum field theories can be applied
without the assumption of integrability. It was started with the analysis of
bulk perturbation \cite{KIM1,KIM2} with Neumann boundary condition, then extended
to boundary perturbations in \cite{NT}-\cite{AbCo}. Most of these papers deal
with comparing exact results, obtained in the aforementioned way for the reflection
matrices in integrable theories, on one hand, and perturbative results on the
other. They defined the reflection matrix \( R(k) \) through the asymptotic
behaviour of the two point function of the field \( \Phi  \) - creating the
particles - far away from the boundary:
\begin{equation}
\label{pertR}
\langle 0|T(\Phi (x,t)\Phi (x^{\prime },t^{\prime }))|0\rangle =\int \frac{d\omega }{2\pi }\frac{e^{-i\omega (t-t^{\prime })}}{2k(\omega )}(e^{ik(\omega )|x-x^{\prime }|}+R(k)e^{-ik(\omega )(x+x^{\prime })})+\dots 
\end{equation}
 No detailed justification has been given for this quantity being the same as
the axiomatic \( R \)-matrix.

In this paper we develop the asymptotic theory for a scalar field with the most
general bulk and boundary self-interaction. We derive the boundary reduction
formula, which connects the reflection matrix to the two point function. As
a consequence we can the fill the gap mentioned above, that is we are able to
connect the \( R \)-matrix of the axiomatic theory to (\ref{pertR}). 

The paper is organized as follows: We apply the canonical quantization procedure
to the free theory, in which case the boundary condition is Neumann. The interacting
theory is defined by means of the adiabatic hypothesis. Asymptotic states and
reflection matrices are introduced and the simplest physical process of one
incoming particle is demonstrated. As the main result we derive the boundary
reduction formula. Having developed the boundary perturbation theory we are
able to connect the earlier definitions of the \( R \)-matrix, and finally
we conclude on their equivalence. The brief explanation of the Feynman rules
is placed in Appendix A, while the structure of the two point function is analysed
in Appendix B.

\subsection*{The free theory}

The system we are dealing with contains a real scalar field \( \Phi (x,t) \),
living on the half space \( x\leq 0 \). The bulk and boundary interactions
are described by the action:

\begin{equation}
\label{S}
S=\int ^{\infty }_{-\infty }dt\int _{-\infty }^{0}dx\left[ \frac{1}{2}((\partial _{t}\Phi )^{2}-(\partial _{x}\Phi )^{2}-m^{2}\Phi ^{2})-V(\Phi )\right] -\int ^{\infty }_{-\infty }dtU(\Phi (0,t))\qquad .
\end{equation}
The free theory can be obtained by switching off the bulk and the boundary interactions:
\( V(\Phi )=U(\Phi )=0 \). The equation of motion is the usual bulk free equation,
the boundary condition is, however, the Neumann one: 
\[
\left( \square +m^{2}\right) \Phi (x,t)=0\quad ;\qquad \partial _{x}\Phi (x,t)|_{x=0}=0\quad .\]
In solving these equations by Fourier transformation we have to use the complete
system of functions with this boundary condition:
\[
\Phi (x,t)=\int _{-\infty }^{\infty }\frac{dk}{2\pi }\cos (kx)\tilde{\Phi }(k,t)\quad ;\quad \tilde{\Phi }(k,t)=\tilde{\Phi }(-k,t)\quad .\]
The conjugate momentum also satisfies Neumann boundary condition so the canonical
commutation relation reads as follows:
\[
[\Phi (x,t),\Pi (x^{\prime },t)]=i\delta _{N}(x,x^{\prime })\equiv i\delta (x-x^{\prime })+i\delta (x+x^{\prime })\quad .\]
The creation and annihilation operators which diagonalise the Hamiltonian are
\[
a(k,t)=i\tilde{\Pi }(k,t)+\omega (k)\tilde{\Phi }(k,t)\quad ;\quad a^{+}(k,t)=-i\tilde{\Pi }(k,t)+\omega (k)\tilde{\Phi }(k,t)\; ,\]
where \( \omega (k)=\sqrt{k^{2}+m^{2}} \). Their commutation relations are
\[
[a(k,t),a^{+}(k^{\prime },t)]=2\pi 2\omega (k)\left( \delta (k-k^{\prime })+\delta (k+k^{\prime })\right) \quad .\]
 Normal ordering is defined as usual: creation operators \( a^{+}(k,t) \) are
to the left of annihilation operators \( a(k^{\prime },t) \). Since the normal
ordered Hamiltonian is 
\[
H=\frac{1}{2}\int _{-\infty }^{\infty }d\tilde{k}\omega (k)a^{+}(k,t)a(k,t)\quad ;\quad d\tilde{k}=\frac{dk}{2\pi 2\omega (k)}\]
 the time dependence can be determined exactly: \( a^{+}(k,t)=a^{+}(k)e^{i\omega (k)t} \)
and \( a(k,t)=a(k)e^{-i\omega (k)t} \). Putting it back into the expansion
of \( \Phi  \) gives rise to 
\begin{equation}
\label{afrel}
\Phi (x,t)=\int ^{\infty }_{-\infty }d\tilde{k}\cos (kx)(a^{+}(k)e^{i\omega (k)t}+a(k)e^{-i\omega (k)t})\; .
\end{equation}
The Fock Hilbert space \( {\cal H} \) can be built up by acting by the creation
operators on the vacuum: 
\[
a(k)|0\rangle =0\quad ;\quad \forall k\]
 
\[
|k_{1},k_{2},\dots ,k_{n}\rangle =a^{+}(k_{1})a^{+}(k_{2})\dots a^{+}(k_{n})|0\rangle \quad ;\quad k_{1}\geq k_{2}\geq \dots \geq k_{n}\geq 0\quad .\]
Note that in labelling the states, \( k \) is always positive.\footnote{%
actually \( k \) provides only a different parametrisation of the energy by
the relation \( k=\sqrt{\omega ^{2}-m^{2}} \), since in the presence of a boundary
the momentum is not conserved.
} For technical reasons in some formulas we also allow \( k \) to take negative
values, but we always mean a symmetric extension, that is \( a(k)=a(-k) \).
The vacuum expectation value of the time ordered product

\begin{equation}
\label{g2free}
\langle 0|T(\Phi (x,t)\Phi (x^{\prime },t^{\prime }))|0\rangle =\int _{-\infty }^{\infty }\frac{d^{2}k}{(2\pi )^{2}}\frac{ie^{-ik_{0}(t-t^{\prime })}}{k^{2}-m^{2}+i\epsilon }(e^{ik_{1}(x-x^{\prime })}+e^{ik_{1}(x+x^{\prime })})
\end{equation}
solves the inhomogeneous equation:
\[
\left( \square +m^{2}\right) \langle 0|T(\Phi (x,t)\Phi (x^{\prime },t^{\prime }))|0\rangle =-i\delta _{N}(x,x^{\prime })\delta (t-t^{\prime })\; .\]
Besides the usual bulk propagator, which describes how the field propagates
from \( (x,t) \) to \( (x^{\prime },t^{\prime }) \), (\ref{g2free}) also
contains another contribution, which can be interpreted as a bulk propagation
of the field from \( (-x,t) \) to \( (x^{\prime },t^{\prime }) \). Thus the
free boundary theory (Neumann boundary condition) can be realized by the mirror
trick: We compute every quantity in the usual bulk theory, but any time we insert
a field at \( (x,t) \) we insert the same type of field also at the mirror
point \( (-x,t) \). Since the interacting theory is defined in terms of the
free quantities, (in the calculations we use the free propagator) we have the
following interpretational consequence: The particles interact not only with
themselves but also with their mirror partners.

\subsection*{Interacting theory, asymptotic states}

Non trivial interaction is described by (\ref{S}) when \( U(\Phi ) \),\( V(\Phi ) \)
or both are non zero. To handle this case we use the adiabatic hypothesis. That
is the interaction is switched on adiabatically in the remote past and switched
off in the remote future. Moreover, we also suppose that the particle spectrum
does not change during this adiabatic procedure: Only the masses are renormalised.
In a real scattering experiment the prepared state is one of the free theory
(\emph{in} state) and the detected state is also a free state (\emph{out} state).
Both the \emph{in} and the \emph{out} states provide a basis for the Hilbert
space \( {\cal H} \), and the \( R \)-matrix is a unitary transformation connecting
the two: 
\[
|final\rangle _{out}=R|initial\rangle _{in}\quad .\]
The unitarity of the \( R \)-matrix expresses the fact that the transition
probabilities sum up to one.

The simplest process we can imagine is in which a single particle travels toward
the boundary, reflects on it and then returns as a multi-particle configuration.
The naive bulk analogue of this process is trivial since asymptotic particles
are stable by assumption. In the presence of the boundary we can interpret this
process in terms of the mirror transformation. In this language the \emph{in}
state contains not only the incoming particle but also its mirror image w.r.t.
the boundary \( x=0 \). The scattering of the incoming particle on the boundary
has a contribution describing its scattering on its mirror image, a process
analogous to the two particle scattering in the bulk. To be concrete: In the
initial state of this process we have a wave-packet of the form 
\begin{equation}
\label{in}
|initial\rangle _{in}=\int _{-\infty }^{\infty }d\tilde{k}^{\prime }f(k^{\prime })|k^{\prime }\rangle _{in}\quad ,
\end{equation}
 where we suppose that \( f(k^{\prime }) \) is well localised around \( k \).
The space time dependence of such a configuration is 
\[
\tilde{f}(x,t)=\int _{-\infty }^{\infty }\tilde{d}k^{\prime }f(k^{\prime })\cos (k^{\prime }x)e^{-i\omega (k^{\prime })t}\quad .\]
It describes a wave packet travelling with momentum \( k \) towards the boundary.
It also contains however, the mirror image of the packet which is on the other
side of the wall (so is not in the real space-time) and travels with momentum
\( -k \) as shown on the figure: 

\vspace{0.3cm}
{\par\centering \resizebox*{9cm}{!}{\includegraphics{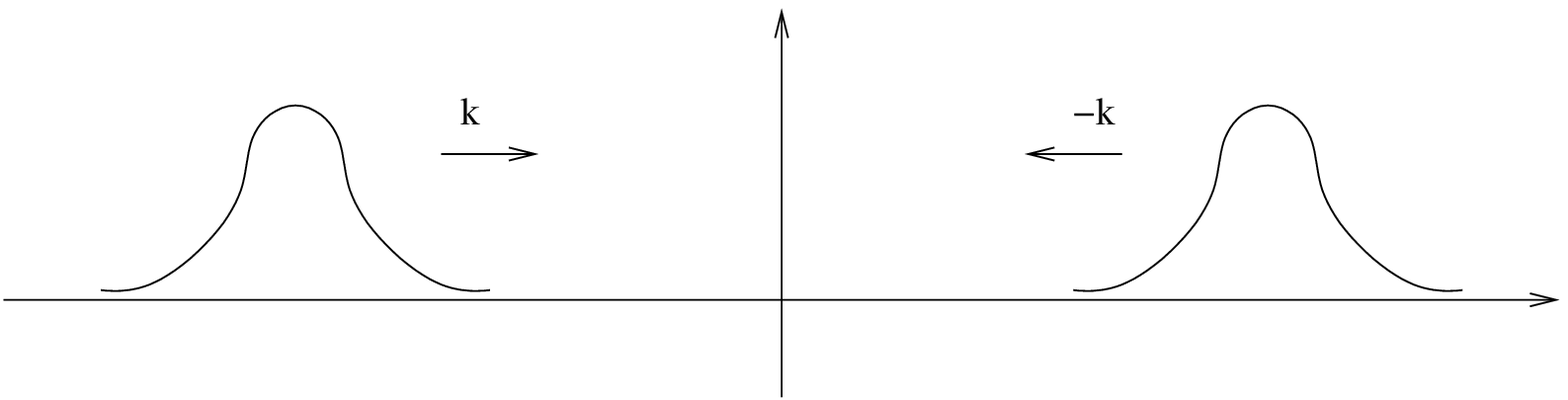}} \par}
\vspace{0.3cm}

If there is no interaction (free case) then this state (\ref{in}) is the eigenstate
of the free Hamiltonian. Since the time evolution is trivial the picture in
the remote future looks like 

\vspace{0.3cm}
{\par\centering \resizebox*{9cm}{!}{\includegraphics{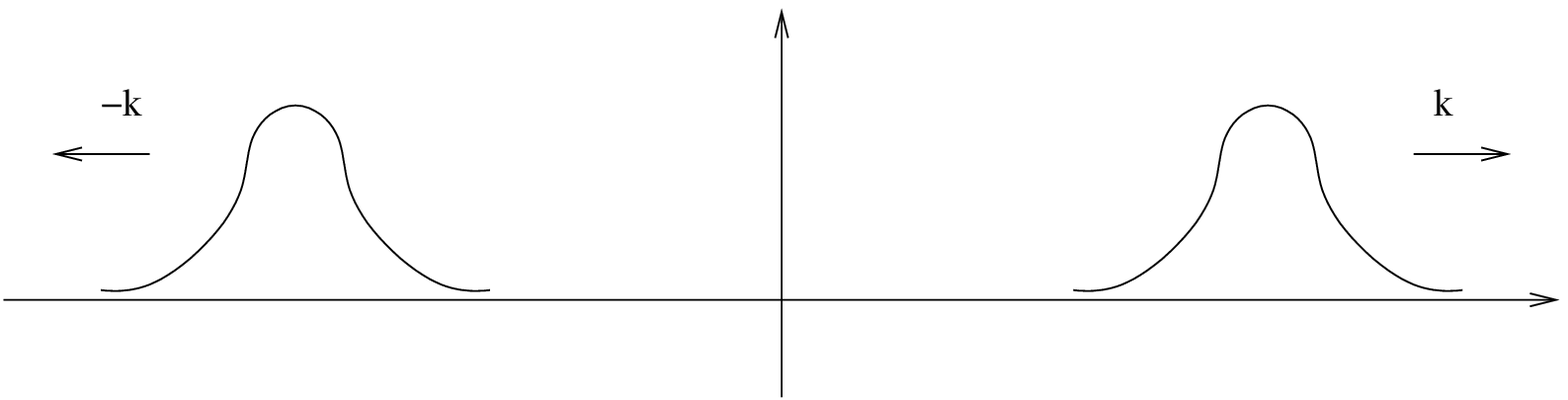}} \par}
\vspace{0.3cm}

Now the real and reflected particle travels away from the boundary with momentum
\( |-k| \) and the mirror image with momentum \( k \).

In the interacting case the final state may contain particles (or just one particle
in the integrable case) travelling backward from the boundary. This coincides
with the idea of \cite{GZ} where the \emph{in} state contains a particle with
rapidity \( \theta  \) while the out state with rapidity \( -\theta  \).

\subsection*{Boundary reduction formula}

We are interested in the case when both the \emph{in} and \emph{out} state contain
a single particle with definite energy. The energy conservation can be factored
out:
\[
_{in}\langle k^{\prime }|R|k\rangle _{in}=2\pi (\delta (k-k^{\prime })+\delta (k+k^{\prime }))\omega (k){\cal R}(|k|)\quad .\]
Our aim is to make correspondence with the other definitions of the reflection
matrix. For this reason we express the reflection matrix \( {\cal R} \) in
terms of the correlation functions. In the bulk theory this is done using the
reduction formula \cite{IZ}. In the following we derive an analogous formula
for boundary theories. The steps of the derivation are similar to the ones in
\cite{IZ}.

The \emph{in} field is a free field so we can use the decomposition (\ref{afrel}).
The inverse of this relation is
\begin{eqnarray}
a_{in}(k) & = & 2i\int _{-\infty }^{0}dx\cos (kx)e^{i\omega (k)t}{\mathop {\partial }^{\leftrightarrow }}_{t}\Phi _{in}(x,t)\label{aa+} \\
a_{in}^{+}(k) & = & -2i\int _{-\infty }^{0}dx\cos (kx)e^{-i\omega (k)t}{\mathop {\partial }^{\leftrightarrow }}_{t}\Phi _{in}(x,t)\; .\nonumber 
\end{eqnarray}
 Using the definition of the \emph{in} state we have
\begin{equation}
\label{matel}
_{out}\langle p_{1},\dots ,p_{k}|q_{1},\dots ,q_{l}\rangle _{in}=:\langle \, \rangle =_{out}\langle p_{1},\dots ,p_{k}|a_{in}^{+}(q_{1})|q_{2},\dots ,q_{l}\rangle _{in}\; .
\end{equation}
Now apply formulas (\ref{aa+}) to obtain
\[
\langle \, \rangle =-2i\int _{-\infty }^{0}dx\cos (q_{1}x)e^{-i\omega (q_{1})t}{\mathop {\partial }^{\leftrightarrow }}_{t}\, _{out}\langle p_{1},\dots ,p_{k}|\Phi _{in}(x,t)|q_{2},\dots ,q_{l}\rangle _{in}\; .\]
We suppose that the \emph{in} field can be expressed in terms of the interacting
field as \( \Phi (x,t)\to Z^{1/2}\Phi _{in}(x,t) \) as \( t\to -\infty  \).
As a consequence 
\[
\langle \, \rangle =-i\lim _{t\to -\infty }Z^{-1/2}2\int _{-\infty }^{0}dx\cos (q_{1}x)e^{-i\omega (q_{1})t}{\mathop {\partial }^{\leftrightarrow }}_{t}\, _{out}\langle p_{1},\dots ,p_{k}|\Phi (x,t)|q_{2},\dots ,q_{l}\rangle _{in}\; .\]
Since \( \displaystyle \mathop {\lim }_{t\to \infty }\Phi (x,t)=\mathop {\lim }_{t\to \infty }Z^{1/2}\Phi _{out}(x,t) \)
we also have
\begin{eqnarray*}
\langle \, \rangle  & = & _{out}\langle p_{1},\dots ,p_{k}|a_{out}^{+}(q_{1})|q_{2},\dots ,q_{l}\rangle _{in}+\\
 &  & iZ^{-1/2}2\int _{-\infty }^{0}dx\int ^{\infty }_{-\infty }dt\partial _{t}\{\cos (q_{1}x)e^{-i\omega (q_{1})t}{\mathop {\partial }^{\leftrightarrow }}_{t}\, _{out}\langle p_{1},\dots ,p_{k}|\Phi (x,t)|q_{2},\dots ,q_{l}\rangle _{in}\}\; ,
\end{eqnarray*}
from which the connected part is 
\[
iZ^{-1/2}2\int _{-\infty }^{0}\! \! dx\int ^{\infty }_{-\infty }\! \! dte^{-i\omega (q_{1})t}\{\cos (q_{1}x)\langle out|\partial _{t}^{2}\Phi (x,t)|in\rangle +\langle out|\Phi (x,t)|in\rangle (-\partial _{x}^{2}+m^{2})\cos (q_{1}x)\}\; ,\]
where \( \langle out| \), (\( |in\rangle  \)) is the shorthand form for \( \langle p_{1},\dots ,p_{k}| \),
(\( |q_{2},\dots ,q_{l}\rangle  \)), respectively. Performing the partial integration,
(which is legitimate if momenta are smeared with wave packets of the form of
(\ref{in})), we have to be careful to keep the surface term. The connected
part turns out to be
\[
iZ^{-1/2}2\int d^{2}xe^{-i\omega (q_{1})t}\cos (q_{1}x)\{\square +m^{2}+\delta (x)\partial _{x}\}\langle out|\Phi (x,t)|in\rangle \; ,\]
where \( \int d^{2}x=\int _{-\infty }^{0}dx\int _{-\infty }^{\infty }dt \)
is the integral over the entire physical space-time. This is the first stage
of the reduction formula. In the second step we eliminate an outgoing particle.
The derivation straightforwardly follows the combination of the previous computation
and the usual bulk derivation. The connected part of the result is
\begin{eqnarray*}
_{out}\langle p_{1},\dots ,p_{k}|q_{1},\dots ,q_{l}\rangle _{in} & = & -4Z^{-1}\int d^{2}xd^{2}x^{\prime }e^{i(\omega (p_{1})t^{'}-\omega (q_{1})t)}\cos (q_{1}x)\cos (p_{1}x^{'})\\
 &  & \hspace {-4cm}\Big \{\square +m^{2}+\delta (x)\partial _{x}\Big \}\Big \{\square ^{\prime }+m^{2}+\delta (x^{'})\partial _{x^{'}}\Big \}\langle p_{2},\dots ,p_{n}|T(\Phi (x,t)\Phi (x^{'},t^{'}))|in\rangle \; .
\end{eqnarray*}
Iterating the steps above the general matrix element (\ref{matel}) can be expressed
in terms of the \( k+l \) point function. 

In particular for the reflection matrix we have
\begin{eqnarray}
 &  & _{out}<k^{'}|k>_{in}-_{out}<k^{'}|k>_{out}=2\pi (\delta (k-k^{\prime })+\delta (k+k^{\prime }))\omega (k)({\cal R}(|k|)-1)=\label{br2} \\
 &  & -4Z^{-1}\int d^{2}xd^{2}x^{\prime }e^{i(\omega (p_{1})t^{'}-\omega (q_{1})t)}\Big \{\square +m^{2}+\delta (x)\partial _{x}\Big \}\Big \{\square ^{\prime }+m^{2}+\delta (x^{'})\partial _{x^{'}}\Big \}G(x,x^{'},t-t^{'})\, ,\nonumber 
\end{eqnarray}
where
\begin{equation}
\label{tpf}
G(x,x^{'},t-t^{'})=\langle 0|T(\Phi (x,t)\Phi (x^{'},t^{'}))|0\rangle \quad .
\end{equation}

\subsection*{Perturbation theory}

Let us turn to the description of the interacting theory as a perturbation of
the free one. In doing so we use the interaction representation. That is the
time evolution operator is given by the time ordered \( (T) \) product as
\[
U(t)=T\exp \left\{ -i\int _{-\infty }^{t}dt^{'}H_{int}(t^{'})\right\} \quad .\]
This Hamiltonian contains the \emph{in} fields and acts on the \emph{in} Hilbert
space by construction. Clearly the \( R \)-matrix can be expressed as 
\[
R=U(\infty )=T\exp \left\{ -i\int _{-\infty }^{\infty }dt^{'}H_{int}(t^{'})\right\} \; ,\]
which also gives a direct calculation of this quantity. The interacting field
is built up from the free field as 
\[
\Phi (x,t)=U^{-1}(t)\Phi _{in}(x,t)U(t)\quad .\]
Putting this expression into the two-point function (\ref{tpf}) and using the
usual heuristic derivation we obtain
\[
\langle 0|T(\Phi (x,t)\Phi (x^{'},t^{'}))|0\rangle =\frac{\langle 0|T(\Phi _{in}(x,t)\Phi _{in}(x^{'},t^{'})\exp \left\{ i\int d^{2}x{\cal L}_{int}[\Phi _{in}(x,t)]\right\} )|0\rangle }{\langle 0|T(\exp \left\{ i\int d^{2}x{\cal L}_{int}[\Phi _{in}(x,t)]\right\} )|0\rangle }\quad .\]
Making an expansion of the exponential we arrive at the perturbative series
\begin{equation}
\label{pert}
=\frac{\sum _{n=0}^{\infty }\frac{i^{n}}{n!}\langle 0|T(\Phi _{in}(x,t)\Phi _{in}(x^{'},t^{'})\int d^{2}x_{1}{\cal L}_{int}[\Phi _{in}(x_{1},t_{1})]\dots \int d^{2}x_{n}{\cal L}_{int}[\Phi _{in}(x_{n},t_{n})])|0\rangle }{\sum _{n=0}^{\infty }\frac{i^{n}}{n!}\langle 0|T(\int d^{2}x_{1}{\cal L}_{int}[\Phi _{in}(x_{1},t_{1})]\dots \int d^{2}x_{n}{\cal L}_{int}[\Phi _{in}(x_{n},t_{n})])|0\rangle }\quad .
\end{equation}
 In computing the vacuum expectation values of the product of the fields we
can use Wick's theorem. The results are encoded in the Feynman rules which are
given in Appendix A. 

From careful analysis of the perturbative series (see Appendix B) one can deduce
that the momentum space Green function has the following form:
\begin{equation}
\label{rppo}
G(p,p^{\prime },\omega )=2\pi (\delta (p+p^{\prime })+\delta (p-p^{\prime }))G_{bulk}(p,\omega )+G_{bulk}(p,\omega )B(p,p^{\prime },\omega )G_{bulk}(p^{\prime },\omega )\quad .
\end{equation}
 Here \( G_{bulk}(p,\omega ) \) is the propagator of the bulk theory, which
in terms of the spectral function \( \sigma (m^{2}) \) has the usual Källen-Lehmann
representation \cite{IZ}:
\begin{equation}
\label{KL}
G_{bulk}(p,\omega )=\frac{iZ}{\omega ^{2}-p^{2}-m^{2}+i\epsilon }+\int _{4m^{2}}^{\infty }dm^{\prime 2}\frac{i\sigma (m^{\prime 2})}{\omega ^{2}-p^{2}-m^{\prime 2}+i\epsilon }\quad .
\end{equation}
We also have the decomposition
\begin{equation}
\label{b123}
B(p,p^{\prime },\omega )=B_{1}(p,p^{\prime },\omega )+B_{2}(p,\omega )+B_{2}(p^{\prime },\omega )+B_{3}(\omega ).
\end{equation}
 The interpretation of the terms in (\ref{rppo}) is the following: The first
term describes the propagation in the presence of the boundary without hitting
the boundary. In the second term \( G_{bulk}(p,\omega ) \) is the propagator
to the boundary, \( B(p,p^{\prime },\omega ) \) is the reflection on the boundary,
while \( G_{bulk}(p^{\prime },\omega ) \) describes the propagation back from
the boundary. In the reflection matrix \( B_{1}(p,p^{\prime },\omega ) \) really
depends on both momenta and comes from the purely bulk interactions, \( B_{3}(\omega ) \)
is the purely boundary contribution and \( B_{2} \) represents the cross terms. 

Now we are able to relate the two different definitions of the \( R \)-matrix.
Performing both inverse Fourier transformations of (\ref{rppo}) in the momentum
variables, but keeping only the contributions of the poles of the first term
in propagators (\ref{KL}):
\[
G(x,x^{\prime },t-t^{\prime })=\int \frac{d\omega }{2\pi }\frac{Ze^{-i\omega (t-t^{\prime })}}{2k(\omega )}(e^{ik(\omega )|x-x^{\prime }|}+(1+\frac{ZB(k(\omega ),k(\omega ),\omega )}{2k(\omega )})e^{-ik(\omega )(x+x^{\prime })})\; ,\]
 where \( k(\omega )=\sqrt{\omega ^{2}-m^{2}} \). Comparing the reflected wave
with the unreflected one the reflection matrix was defined to be 
\[
R(k)=1+\frac{ZB(k(\omega ),k(\omega ),\omega )}{2k(\omega )}.\]

We will recover the same result from our boundary reduction formula (\ref{br2}).
First we recall that the reduction formula describes the way how the matrix
elements are related to the correlation functions. Considering the correlation
functions in momentum space the operator \( \square +m^{2} \) gives a factor
of \( -k^{2}+m^{2} \) for each external leg. The space-time integrations, as
inverse Fourier transformations, put all the momenta on shell. Since all of
the outer legs in the correlation functions are dressed up in the perturbation
theory to contain the exact bulk propagators (\ref{KL}) with poles of the form
\( \frac{iZ}{k^{2}-m^{2}} \), the reduction formula merely amputates the legs
and gives the residue of this multi-pole pole. In the boundary case we have
an analogous interpretation. Similarly to the bulk case the momentum conserving
part of (\ref{rppo}) does not give any contribution to the \( R \)-matrix
so it is enough to consider the other term. A careful analysis shows that \( (\square +m^{2}) \)
is the operator which amputates the legs starting with a bulk vertex, while
\( \delta (x)\partial _{x} \) is responsible for amputation of the legs starting
with a boundary vertex. As a consequence the \( (\square +m^{2})(\square ^{\prime }+m^{2}) \)
term gives \( B_{1}(k,k^{\prime },\omega (k)) \), the terms \( (\square +m^{2})\delta (x^{\prime })\partial _{x^{\prime }} \)
and \( (\square ^{\prime }+m^{2})\delta (x)\partial _{x} \) together give \( B_{2}(k,\omega (k))+B_{2}(k^{\prime },\omega (k)) \),
finally \( \delta (x)\partial _{x}\delta (x^{\prime })\partial _{x^{\prime }} \)
gives \( B_{3}(\omega (k)) \). We also have an overall factor \( 2\pi Z\delta (\omega (k)-\omega (k^{\prime })) \)
expressing energy conservation. Collecting all these terms and using the identity
\( 2\pi \delta (\omega (k)-\omega (k^{\prime }))=\frac{\omega (k)}{k}2\pi (\delta (k-k^{\prime })+\delta (k+k^{\prime })) \),
we obtain that 
\[
\mathcal{R}(k)=1+\frac{ZB(k,k,\omega (k))}{2k}\quad ,\]
which shows that the reflection factor defined with use of the asymptotic states
and the one defined using the two-point function are identical.

\subsection*{Conclusion}

The boundary reduction formula, derived in the paper for a boson with the most
general (possibly non-integrable) boundary and bulk self-interaction, showed
the equivalence of the previously used definitions for the \( R \)-matrix,
and supports the correctness of the formalism. This formulation enables one
to derive the main properties of the \( R \)-matrix such as analyticity, unitarity,
crossing symmetry and analyse its pole structure directly without referring
to the crossed channel picture used in \cite{GZ}. The analysis of the perturbative
series, Landau equations, Cutkosky rules, the derivation of the boundary Coleman-Thun
mechanism and of the analyticity properties of the \( R \)-matrix are the subjects
of our next paper.

\subsubsection*{Acknowledgments}

We thank L. Palla, Z. Horváth, Gy. Pócsik and F. Csikor for valuable discussions.
Z.B. and G.B. were supported by Bolyai Fellowships and G. T. by a Magyary Fellowship.
This research was supported in part by FKFP 0043/2001, OTKA 034512, 037674,
034299.

\subsection*{Appendix A}

The Feynman rules of the boundary theory can be obtained by expanding the interaction
Lagrangean as 
\begin{equation}
\label{lpert}
{\cal L}_{int}(x,t)=\sum _{N=0}^{\infty }\frac{\alpha _{N}}{N!}\Phi (x,t)^{N}+\delta (x)\sum _{M=0}^{\infty }\frac{\beta _{M}}{M!}\Phi (0,t)^{M}
\end{equation}
and putting (\ref{lpert}) into (\ref{pert}). In contrast to the perturbative
approaches developed so far we formulate the resulting Feynman rules in momentum
space. That is we compute 
\[
{\tilde{G}}(p_{1},\omega _{1},\dots p_{n},\omega _{n})=\int _{-\infty }^{\infty }dx_{1}\dots \int _{-\infty }^{\infty }dx_{n}\int _{-\infty }^{\infty }dt_{1}\dots \int _{-\infty }^{\infty }dt_{n}e^{i\sum _{j=1}^{n}(p_{j}x_{j}-\omega _{j}t_{j})}G(x_{1},t_{1}\dots x_{n},t_{n})\]
 according to the Feynman rules: 

\begin{itemize}
\item Draw all possible oriented graphs with \( n \) external legs such that the
external legs point inside the graph. The lines (both external and internal)
can be of two types: straight and dashed. The vertices can be two types: black
and white. Label the external legs by the 2-momenta \( (p_{1},\omega _{1})\dots (p_{n},\omega _{n}) \),
and the internal lines by new momentum variables \( (k^{1},k^{0}) \).
\end{itemize}
\vspace{0.3cm}
{\par\centering \resizebox*{8cm}{!}{\includegraphics{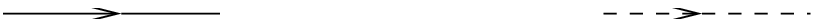}} \par}
\vspace{0.3cm}

\begin{itemize}
\item To a black vertex with \( N \) incident lines attache the contribution 
\begin{equation}
i\alpha _{N}2\pi \delta (\sum _{i\in in}k_{i}^{0}-\sum _{i\in out}k_{i}^{0})\pi \delta (\sum _{i\in in}^{str}k_{i}^{1}-\sum _{i\in in}^{dsh}k_{i}^{1}-\sum _{i\in out}k_{i}^{1})
\end{equation}

\end{itemize}
\vspace{0.3cm}
{\par\centering \resizebox*{!}{1.5cm}{\includegraphics{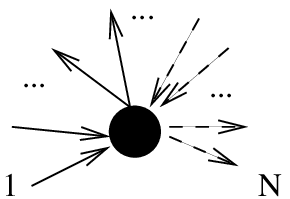}} \par}
\vspace{0.3cm}

\begin{itemize}
\item To a white vertex with \( M \) incident lines attach the contribution 
\begin{equation}
i\beta _{M}2\pi \delta (\sum _{i\in in}k_{i}^{0}-\sum _{i\in out}k_{i}^{0})
\end{equation}

\end{itemize}
\vspace{0.3cm}
{\par\centering \resizebox*{!}{1.5cm}{\includegraphics{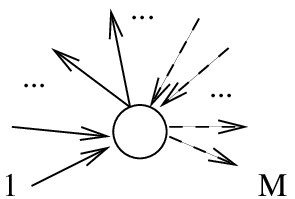}} \par}
\vspace{0.3cm}

\begin{itemize}
\item To a line of any type labelled by \( (k^{1},k^{0}) \) attache the contribution
\begin{equation}
\frac{i}{k^{2}-m^{2}+i\epsilon }
\end{equation}
 If the line is internal then integrate over \( k \): \( \int \frac{d^{2}k}{(2\pi )^{2}} \).
\item Sum over all topologically distinct diagrams. 
\end{itemize}
The notation \( i\in in/out \) means that the line labelled by \( (k_{i}^{1},k_{i}^{0}) \)
is in-coming/out-going at the given vertex. The label \( str/dsh \) on the
sum means that the summation goes only over straight/dashed lines.

These Feynman rules need some explanation. The derivation is fairly standard:
One derives the graph rules first in coordinate space. This is done by writing
\( G(x_{1},t_{1}\dots x_{n},t_{n}) \) as a perturbative series. Each term of
the series is represented by a graph by associating the various parts of the
contribution to the vertices and lines of the graph. In particular, there is
a space-time integration associated to each vertex. According to the two terms
in (\ref{lpert}), there are two kinds of vertices: black (bulk) ones corresponding
to the first term and white (boundary) ones corresponding to the second term.
Owing to the presence of the factor \( \delta (x) \) in the second term of
(\ref{lpert}) at the boundary vertices the space integration can be performed.
There is only a single type of graph lines at this stage.

Converting to momentum space (Fourier transformation of \( G(x_{1},t_{1}\dots x_{n},t_{n}) \)),
all the remaining space-time integrations can be performed resulting \( \delta  \)
functions at each vertex on the momenta of the lines incident to the vertex.
At the boundary vertex there is only a time integration left, hence we obtain
a \( \delta  \) function involving only the time-like component of the momenta.
At the bulk vertex there are both time and space integrations resulting in two
\( \delta  \) functions. The presence of both \( x_{i}-x_{j} \) and \( x_{i}+x_{j} \)
dependent terms in the free propagator (\ref{g2free}) implies that we have
a sum of \( \delta  \) functions on the space-like component of the momenta.
The \( \delta  \)'s in this sum differ in the signs of some momenta in the
argument. The two types of lines are introduced in order to associate different
graphs to the individual terms: a line is straight if it corresponds to the
term coming from the \( x_{i}-x_{j} \) dependent part of the free propagator
associated to the given line, and dashed if it comes from the \( x_{i}+x_{j} \)
dependent part.

\subsection*{Appendix B }

In this appendix we analyse the properties of the two point function in momentum
space. A systematic investigation of the correlation functions can be achieved
by generalising the parametric representation (\cite{IZ} section 6.2.3 ) to
the boundary case. Since the introduction of all the machinery is quite lengthy
and we need the result only for the two point function, instead of the detailed
presentation we summarise how the formula (\ref{rppo}) can be obtained by the
direct analysis of the perturbative series. 

Energy is conserved at each vertex so we have
\[
\tilde{G}(p_{1},\omega _{1},p_{2},\omega _{2})=2\pi \delta (\omega _{1}-\omega _{2})G(p_{1},p_{2},\omega _{1})\quad .\]
Momentum is not conserved in general. There are graphs whose contribution spoils
momentum conservation: such is any graph containing a boundary vertex. On the
other hand a graph with only bulk vertices and straight lines gives rise to
a contribution proportional to \( \delta (p_{1}-p_{2}) \). Furthermore, the
contribution is proportional to the contribution of the same graph computed
in the bulk theory. This can be verified by comparing the appropriate graph
rules with the bulk theory rules, and one finds that the only difference is
a factor \( 1/2 \) in the bulk vertex contribution. That is the contribution
of a graph with \( N_{b} \) bulk vertices, no boundary vertices and only straight
lines is \( 2^{-N_{b}} \) times the contribution of the same graph was in the
bulk theory.

On the other hand there is a symmetry operation on the graphs that leaves the
contribution invariant. Namely, at each bulk vertex we can change the types
of all incident lines and also the signs of the momenta labelling the \emph{out-going}
lines. One checks that this transformation -- performed independently at each
of the \( N_{b} \) bulk vertices -- does not affect the graph contribution
and gives rise to a symmetry factor \( 2^{N_{b}} \). This symmetry factor just
compensates the factor \( 2^{-N_{b}} \). Hence the sum of the contributions
of the graphs with only bulk vertices and straight lines plus the ones related
to them by the symmetry transformation described gives the propagator of the
bulk theory. It is not difficult to see that no other graph gives contribution
respecting momentum conservation.

Summing then the contributions of the graphs with only bulk vertices and straight
\emph{internal} lines (and arbitrary \emph{external} ones) plus the graphs related
to them by symmetry transformation we obtain the ``momentum-preserving'' part
of the propagator 
\[
2\pi \left( \delta (p_{1}+p_{2})+\delta (p_{1}-p_{2})\right) G_{bulk}(p_{1},\omega _{1}).\]

Now we concentrate on the momentum non-preserving part. A diagram which remains
connected when any of its internal line is cut is called one particle irreducible.
Any diagram with two external legs can be built by attaching one particle irreducible
diagrams after each other. For the momentum non-preserving diagrams we can separate
the consecutive momentum preserving one particle irreducible subdiagrams adjacent
to a given external line. This gives subgraphs that are identical to the graphs
in the series of the bulk propagator, so they give factors \( G_{bulk}(p_{1},\omega _{1}) \)
and \( G_{bulk}(p_{2},\omega _{2}) \) at the external lines, respectively.
The contributions of the remaining momentum non-preserving subgraphs are collected
in \( B(p_{1},p_{2},\omega _{1}) \). As a consequence we have the following
form for the propagator: 
\[
G(p_{1},p_{2},\omega _{1})=2\pi (\delta (p_{1}+p_{2})+\delta (p_{1}-p_{2}))G_{bulk}(p_{1},\omega _{1})+G_{bulk}(p_{1},\omega _{1})B(p_{1},p_{2},\omega _{1})G_{bulk}(p_{2},\omega _{1})\quad .\]
 If in a Feynman graph, contributing to \( B(p_{1},p_{2},\omega _{1}) \), both
external lines are incident to a boundary vertex, then it does not depend on
any of the momenta. Its contribution is collected in \( B_{3}(\omega ) \).
Terms depending on one momentum only are collected in \( B_{2}(p_{1},\omega _{1}) \),
these are the graphs in which one of the external lines ends in a boundary vertex.
The contribution of a diagram starting with bulk vertices on both ends depends
both on \( p_{1} \) and \( p_{2} \). Such contributions are collected in \( B_{1}(p_{1},p_{2},\omega _{1}) \).
Summing up all these terms we have (\ref{b123}).

\end{document}